\documentclass[aps,showpacs,twocolumn,prl,superscriptaddress]{revtex4}
\usepackage{amssymb}
\usepackage{natbib}
\usepackage{amsmath}
\usepackage{amsfonts}
\usepackage{graphicx}
\usepackage{mathrsfs}
\usepackage{dcolumn}
\usepackage{bm}
\usepackage{color}
\usepackage{cancel}
\usepackage{mathbbol}
\usepackage{epsfig}
\usepackage{units}
\usepackage{esint}
\usepackage{indentfirst}
\usepackage{%
 graphicx,
 hyperref,% add hypertext capabilities
 dcolumn,% Align table columns on decimal point
natbib,%
units,
amsmath
}

\begin{document}

\title{Seasonal Floquet states in a game-driven evolutionary dynamics} 
\author{Olena Tkachenko}
\affiliation{Sumy State University, Rimsky-Korsakov Street 2, 40007 Sumy, Ukraine}
\author{Juzar Thingna}
\affiliation{Institut f\"ur Physik, Universit\"at Augsburg,
Universit\"atsstra\ss e 1, 86159 Augsburg, Germany}
\author{Sergey Denisov}
\affiliation{Nanosystems Initiative Munich, Schellingstr, 4, D-80799 M\"{u}nchen, Germany}
\affiliation{Sumy State University, Rimsky-Korsakov Street 2, 40007 Sumy, Ukraine}
\author{Vasily Zaburdaev}
\affiliation{Max Planck Institute for the Physics of Complex Systems, 
N\"{o}thnitzer Str. 38, D-01187 Dresden, Germany}
\author{Peter H\"anggi}
\affiliation{Institut f\"ur Physik, Universit\"at Augsburg,
Universit\"atsstra\ss e 1, 86159 Augsburg, Germany}
\affiliation{Nanosystems Initiative Munich, Schellingstr, 4, D-80799 M\"{u}nchen, Germany}
%\affiliation{Nanosystems Initiative Munich, Schellingstr, 4, D-80799 Munchen, Germany}

\date{\today}

\pacs{02.50.Le, 87.23.Kg, 05.45.-a}
\begin{abstract}

Mating preferences of many biological species are not constant  but season-dependent. 
Within the framework of evolutionary game theory this can be modeled with two finite opposite-sex 
populations playing against each other following the rules that are
periodically changing. By combining Floquet theory and the concept of 
quasi-stationary distributions, we reveal existence of metastable time-periodic 
states in the evolution  of finite game-driven populations. The evolutionary Floquet states 
correspond to  time-periodic probability flows in the strategy space  
which cannot be resolved within the mean-field framework. The lifetime
of metastable Floquet states increases with the size $N$ of populations so that they
become attractors in the limit $N \rightarrow \infty$. 
\end{abstract}
\maketitle
%----------------------------------------------------------------

\textit{Introduction.} The evolutionary dynamics of an animal group is tied to the reproductive activity
of its members, a complex process which involves  courtship rituals and 
sharing of parental care \cite{sexual}. 
Within the  game theory framework, the sex conflict over parental investment was 
formalized by Dawkins  in his famous  ``Battle of Sexes'' (BoS) \cite{dawkins},
illustrated in Fig. 1. In this game two opposite-sex  members of the group
play against each other. Each player can use two  behavioral strategies. 
Entries in the payoff matrix, $b_{ss'}$, quantify the reward received by
a female which used a  strategy $s \in \{1,2\}$ after she has played against a 
male which used a  strategy $s' \in \{1,2\}$. Entries $a_{s's}$ define the reward of the male.
%The average payoff determines the probability of the player to be selected for reproduction. 
%---------------------------------------------------------------------------------
\begin{figure}[!htbp]
\includegraphics[width=0.45\textwidth]{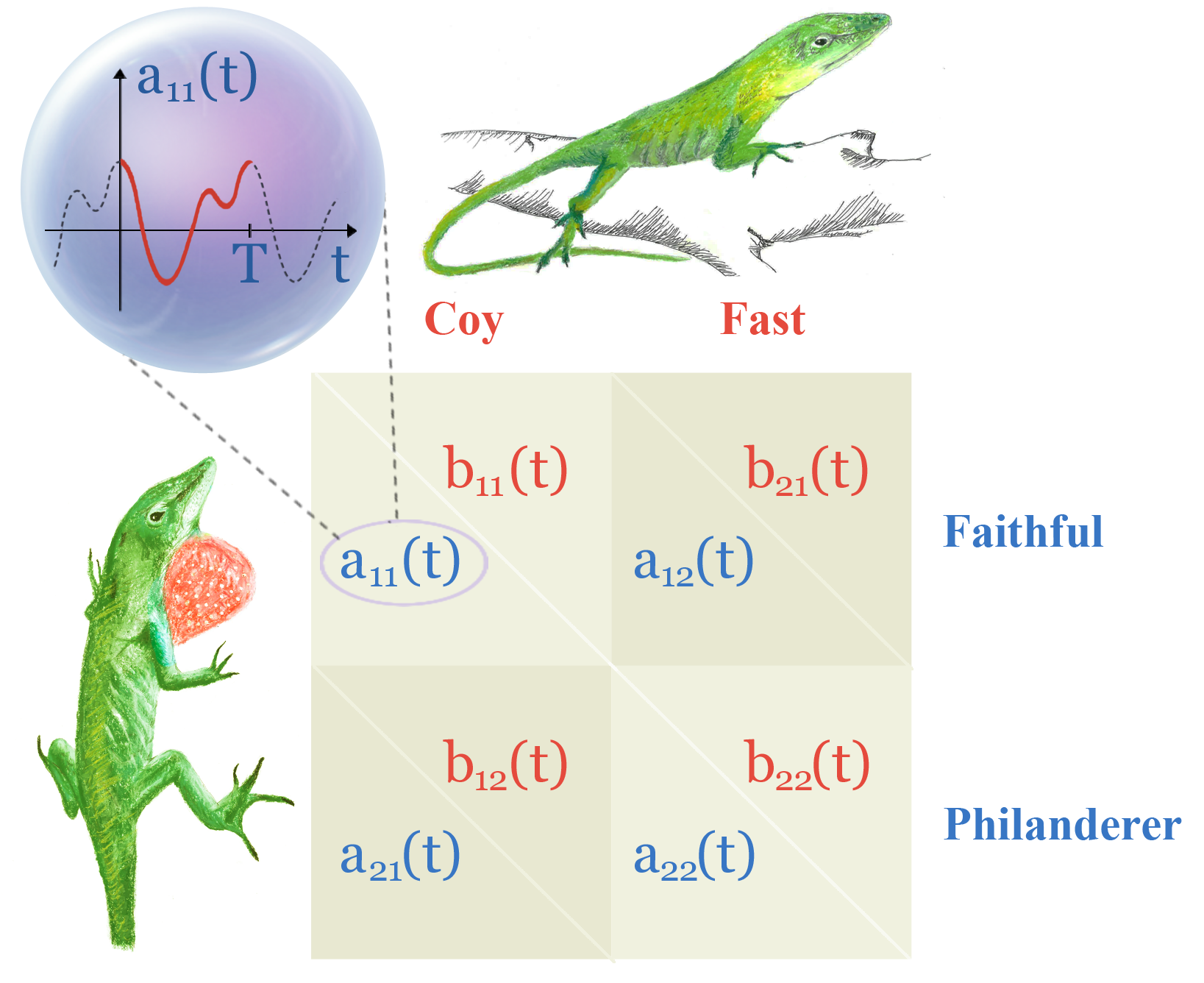}
\caption{(color online) ``Battle of Sexes'' with seasonal variations. 
%It models a mate selection 
%in a population of Carolina anole lizards. 
A female of Carolina anole lizards can be either \textit{coy} and  prefer an 
arduous courtship, to be sure that a mate is ready to contribute to a parental care, or \textit{fast}, 
and thus not being much concerned about parental care of offspring. 
A male can be either \textit{faithful} and ready to assure the female partner,  
by performing a long courtship, that he is a faithful potential husband, or \textit{philanderer} 
and prefer to shorten the courtship stage. 
Depending on the strategies, $s$ ($s'$), played by the female (male), 
the female (male) gets payoff $b_{ss'}$ ($a_{s's}$). 
Both females and males are season-constrained in their strategies and preferences, 
which is modeled via time-periodic modulations of the payoffs.
}\label{Fig:1}
\end{figure}
%---------------------------------------------------------------------------------

A number of observations have shown that mating strategies and preferences of many species
are not constant in time but  season-dependent \cite{generic}. 
For example, courtship srituals of the males of
Carolina anole lizards (\textit{Anolis carolinensis}), as well as mate selection criteria of the females
of the species, are periodically changing during the year  \cite{lizard1}. 
Even the amount of different types of muscle fibers 
that control the vibrations of a red throat fan (dewlap) - which males employ during the courtship - 
is a season dependent characteristic \cite{lizard2}. Currently, there
is no agreement between the ecologists on the
role this seasonal plasticity  plays in  determining 
the evolution direction of the species \cite{srev}.
%The choice of a mate in an animal group can be considered as an optimization problem \cite{pom}.
%When selecting (courting) a mate, a female (male) of a species faces 
%a complex choice problem where benefits of a choice depend on the season 
%and have to be traded off against each other. Within the BoS framework, these 
%situation can be modeled with a game whose payoffs are periodically modulating in time, see Fig. 1.

We address this problem within the BoS framework by allowing the payoffs to periodically vary in time, see Fig. 1.
Our goal is to investigate how these modulations influence 
the game-driven evolutionary dynamics. Here, we first apply the concept of quasi-stationary 
distributions in absorbing Markov chains \cite{darroch} 
to a stochastic evolutionary dynamics of finite populations  and 
define the notion of evolutionary metastable states. 
Then, by employing the Floquet theory \cite{floquet,floquetH}, 
we generalize the notion of metastable states \cite{meta1,meta2, assaf} to periodically 
modulated game-driven evolutionary dynamics. We show that in big but finite populations,
the metastable Floquet states survive over extremely long 
(as compared to the period of modulations) timescales. We argue that, in the limit of infinitely big populations,
these states become attractors while still evading the mean-field
description. 

%These attractors are starkly different 
%from those corresponding to the case of stationary mating strategies, thus underlying the importance of seasonal
%variations  in the evolutionary dynamics of species.

%Floquet metastable states are not only specific to the 
%problem of season-modulated reproduction 
%and can have implications for understanding  non-equilibrium 
%dynamics of finite-size systems in general. 

\textit{Model.} Finite size of animal populations favors a stochastic approach to  evolutionary dynamics. 
Although the convergence  to the deterministic mean-field dynamics is typically guaranteed in 
the limit $N \rightarrow \infty$ \cite{hof2,schlag}, 
the stochastic dynamics of large but finite populations  
can still be very different from the  mean-field picture \cite{traulsen1,traulsen2,traulsen3,frey}.
Here we adapt the game-oriented version of the Moran process \cite{moran}, 
introduced in Ref. \cite{moranNature} and generalized to two-player games 
in Ref. \cite{traulsen1}. 

Players $A$ (males) and $B$ (females) form two 
populations, each one of a fixed size $N$ and with two available strategies, $s = \{1,2\}$, see Fig.~1. 
%Payoffs are specified by four functions, $\{a_{ss'}(t)\}$ and $\{b_{s's}(t)\}$, $s,s' = \{1,2\}$. 
Game payoffs are time-periodic functions, $c_{ss'}(t) = c_{ss'}(t+T)$, $c = \{a,b\}$, and can be 
represented as sums of stationary and zero-mean time-periodic 
components, $c_{ss'}(t) = \bar{c}_{ss'} + \tilde{c}_{ss'}(t)$, $\langle\tilde{c}_{ss'}(t)\rangle_T = 0$. 
The time starting from $t=0$ is incremented by $\triangle t = T/M$ after each round. 
After $M$ rounds the payoffs return to their initial values. 
The state of the populations after the $m$-th round is fully specified by the number of players 
playing the first strategy, $i$ (males) and $j$ (females), $0 \leq i,j 
\leq N$. A detailed description of the corresponding stochastic process is given in Supplemental Material.
It can also be shown that, in 
the limit $N \rightarrow \infty$ \cite{traulsen1}, the dynamics of the variables 
$x = i/N$ and $y = j/N$ is defined by the adjusted replicator equations \cite{generic};
see Refs. \cite{smith,hof2}.

For a finite $N$, the state of the system can be expressed as a $N \times N$ matrix $\mathbf{p}$ with elements $p(i,j)$, 
which are the probabilities to find two populations in the states $i$ and $j$, respectively.
Round-to-round dynamics can be evaluated by multiplying
%\cite{transition} 
the state $\mathbf{p}$ with the transition fourth-order tensor $\mathbf{S}$, 
with elements $S(i,j,i',j')$ \cite{generic}. 
By using the bijection $k = (N - 1)j + i$, we can unfold the probability matrix $p(i,j)$ into 
the vector $\tilde{p}(k)$, $k = 0,...,N^2$, and the tensor $S(i,j,i',j')$ into the 
matrix $\tilde{S}(k,l)$. This reduces the problem to a Markov 
chain \cite{gant}, $\tilde{\textbf{p}}^{m+1} = \tilde{\textbf{S}}^{m}\tilde{\textbf{p}}^{m}$, 
where $m$ is the number of the round to be played. 
The four states $(i = \{0,N\}, j= \{0,N\})$ are absorbing states because the transition rates 
leading out of them equal zero \cite{generic}. The absorbing states are attractors of the dynamics for 
any \textit{finite} $N$, 
and the finite-size fluctuations will eventually drive a population to one of them  \cite{dykman,frey}. 
This would imply a \textit{fixation}, so that 
only one strategy survived in each of now monomorphic populations \cite{traulsen1, frey}.

%---------------------------------------------------------------
\begin{figure*}[t]
\includegraphics[width=0.95\textwidth]{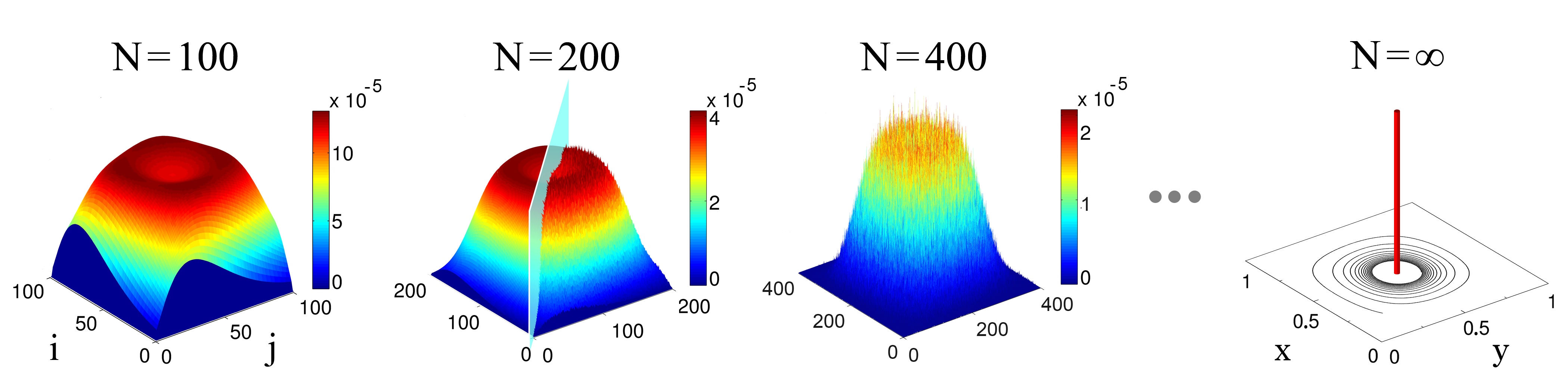}
\caption{(color online) Metastable states of the stationary BoS game. In the mean-filed limit $N = \infty$, 
a trajectory spirals towards a fixed point  
$\left(\frac{1}{2},\frac{1}{2}\right)$, the Nash equilibrium of the game. 
For the finite $N$, metastable states are specified by their 
quasi-stationary probability density functions (pdf's) (3d plots). 
%With the increase of $N$, the functions
%tend to localize at the Nash equilibrium. 
%Although, for any $N$, the mean position $\left(\bar{x}(t), \bar{y}(t)\right)$ coincide with the Nash equilibrium, 
%the  stochastic evolution is governed by the metastable limit cycles located on the crater ridge
%on the pdf's tops. 
For $N=200$ the pdf combines the results of the 
direct diagonalization of the $39~ 601 \times 39~ 601$ matrix $\tilde{\mathbf{Q}}$ (left half of the pdf, 
this procedure was also used to obtain the function for $N=100$) and of the preconditioned stochastic sampling (right part of the pdf, this procedure was also used to obtain 
the function for $N=400$) \cite{generic}. 
The baseline fitness $w = 0.3$ (other parameters are given in the text).}\label{Fig:2}
\end{figure*}
%------------------------------------------------------------------
We are interested in the dynamics before the fixation, 
so we merge the four states into a single absorbing state by summing the corresponding 
incoming rates. The boundary states, $(i = \{0,N\},j\in \{1,\cdots,N-1\})$ and $(i\in \{1,\cdots,N-1\},j=\{0,N\})$, 
can also be merged into this absorbing super-state: 
Once the population gets to the  boundary, it will only move towards 
one of the two nearest absorbing states. 
By labeling the absorbing super-state with index $k = 0$, we end up with a $(L+1) \times (L+1)$ matrix
\begin{eqnarray} 
\tilde{\mathbf{S}}^{m} = \begin{bmatrix}
       1 & \boldsymbol{\varrho_0}^{m} \\
       \mathbf{0} & \tilde{\mathbf{Q}}^{m},      
     \end{bmatrix}
\label{Eq:supermatrix}
\end{eqnarray}
where $L = (N-1)^2$, $\boldsymbol{\varrho}_0^{m}$ 
is a vector of the incoming transition probabilities of 
the absorbing super-state, $\mathbf{0}$ is a $L\times 1$ zero vector, 
and $\tilde{\mathbf{Q}}^{m}$ is a $L \times L$ reduced transition matrix.

With Eq.~(\ref{Eq:supermatrix}), we arrive at  the setup used 
by Darroch and Seneta to formulate the concept of quasi-stationary distributions \cite{darroch}. 
There is the normalized right eigenvector of the reduced 
transition matrix $\tilde{\mathbf{Q}}^{m}$ with the maximum
eigenvalue $\lambda$ \cite{FPT}. By using the inverse bijection, 
we can transform this vector 
%$\tilde{\mathbf{d}}$ 
into a two-dimensional probability density function (pdf), i.e., a state,  $\mathbf{d}$, 
with maximal mean absorption time. 
This state is the most resistant to the wash-out 
by the finite-size fluctuations and it remains near 
invariant, up to a uniform rescaling, under the action of the tensor ${\mathbf{S}}$. 
This is the metastable state of the evolutionary process.

\textit{Stationary case.} As an example, we consider a game with  
payoffs $a_{11}$, $a_{22}$, $b_{12}$ and $b_{21}$ equal $1$, and 
payoffs $-1$ for the rest of strategies  \cite{neumann}. Figure 2 presents 
the numerically obtained metastable states of the game.
We use two methods, the direct diagonalization of the reduced transition matrix,
which is stationary in this case, $\tilde{\mathbf{Q}}^{m} \equiv \tilde{\mathbf{Q}}$,
and preconditioned stochastic sampling \cite{generic}. 
%The latter was performed by 
%launching trajectories from random initial points, 
%uniformly distributed on the $N-1 \times N-1$ 
%grid and then sampling the pdf with only those trajectories which remained unabsorbed after $10\cdot N^2$ rounds.
For $N=200$ we find an agreement between the results of the two methods.
The means of the metastable state, 
\begin{eqnarray}
\bar{x} = \sum_{i,j=1}^{N-1} \frac{i}{N} \cdot d(i,j) ;~\bar{y} = \sum_{i,j=1}^{N-1} \frac{j}{N} \cdot d(i,j),
\label{Eq:mean}
\end{eqnarray}
coincide with the Nash equilibrium \cite{nash} for any $N$. However, 
the actual dynamics is determined by the metastable 
limit cycle encircling the equilibrium  (this could be seen 
by performing short-run stochastic simulations); 
see Fig. 2. Within the Langevin-oriented approach to the 
dynamics of finite populations \cite{traulsen1,traulsen4},
the appearance of the metastable limit cycle can 
be interpreted as a stochastic Hopf bifurcation \cite{arnold} (see also Ref.~\cite{lindner} for another 
interpretation). In the  limit $N \rightarrow \infty$ the cycle collapses to the Nash equilibrium. 
Note, however, that the convergence to this limit is slow, as indicated by the width  of the pdf for $N=400$.

\textit{Case of modulated payoffs}. By adding time-modulations to the model, we find that the mean-field dynamics
does not exhibit substantial changes. For 
the choice $\epsilon(t) = \tilde{a}_{11}(t) = \tilde{b}_{22}(t)= f \cos(\omega t)$ with $\omega = 2\pi/T$ 
(all other payoffs held stationary) we observed a period-one 
limit cycle localized near the Nash equilibrium of the stationary case, 
see Figs.~3(a,b). It collapses to a set of adiabatic Nash 
equlibria, $\left\{x_{NE}(\epsilon) = \frac{2-\epsilon}{4 - \epsilon}, 
y_{NE}(\epsilon) = \frac{2}{4 + \epsilon}\right\}$ in the limit $\omega \rightarrow 0$. 

The dynamics of a finite $N$ population is different.
The  stochastic evolution of a trajectory in the $(i,j)$-space, initiated away from
the absorbing boundary, can be  divided into two stages. 
At first the trajectory relaxes towards a metastable state.
The timescale of this process is defined by the  mixing time $t_{\mathrm{mix}}(N)$ \cite{traulsen5},
which in this case has to be calculated now for the \textit{quasi}-stationary state.
Then the trajectory wiggles  
around the  metastable state until the fluctuations drive it to the absorbing 
boundary. Following the  random-walk approximation, the mean absorption 
time $t_{\mathrm{abs}}(N)$, called ``mean fixation time'' \cite{dawkins,smith} in the evolutionary context,
seemingly should also scale as $N$. 
However, this estimate  neglects the presence of the inner attractive manifold and the fact that 
the noise strength decreases upon approaching the absorbing boundary. 
In fact, the absorption time scales super-linearly with $N$ \cite{meantime}.
The lifetime of the metastable state is restricted to the time interval
$[t_{\mathrm{mix}}(N), t_{\mathrm{abs}}(N)]$, whose length scales 
as $t_{\mathrm{abs}}(N)[1 - t_{\mathrm{mix}}(N)/t_{\mathrm{abs}}(N)] \sim t_{\mathrm{abs}}(N)$.

For $\omega = 0.1$ \cite{size_issue}, the stochastic simulations reveal a metastable state 
which is distinctively different from  
the limit cycle produced by the mean-field equations, see Fig.~3b. There is a conflict between the 
evolution of means, described by the adjusted replicator equations, and the results of 
the stochastic dynamics. The conflict can be resolved with  the concept of the quasi-stationary 
distribution. Namely, the transition matrices, Eq.~(\ref{Eq:supermatrix}), 
are round-specific now and form a set $\{\tilde{\mathbf{S}}^{m}\}$, $m=1,\cdots, M$. 
%(recall that after $M=T/\triangle t$ rounds the periodically modulated payoffs return to their initial value). 
The propagator over the time interval $[0,t]$, $0 < t < T$, is the product 
$\tilde{\mathbf{U}}(t) = \prod_{m'=1}^{M_{t}}\tilde{\mathbf{S}}^{m'}$ with $M_{t}=t/\triangle t$. 
All the propagators, including the period-one propagator $\tilde{\mathbf{U}}(T)$, have the same structure 
as the super-matrix in 
Eq.~(\ref{Eq:supermatrix}). We define the metastable state
$\mathbf{d}(T)$ as the the quasi-stationary distribution of $\tilde{\mathbf{U}}(T)$. 
It is also a \textit{Floquet} state \cite{floquetH} of the reduced propagator 
$\tilde{\mathbf{U}}^{r}(T)$,
which can be obtained by replacing the transition matrices $\tilde{\mathbf{S}}^{m'}$ with the  
matrices $\tilde{\mathbf{Q}}^{m'}$ or by simply cutting
out the first line and column from the matrix $\tilde{\mathbf{U}}(T)$. 
The Floquet state is a time-periodic state, $\mathbf{d}(t + T) = \mathbf{d}(t)$, 
which changes during one period of modulations, see Fig.~4. 
The metastable state $\mathbf{d}(t)$ at any instant of time $t$, $0 < t <T$, 
can be obtained by acting on the state $\mathbf{d}(0)$ with  the reduced propagator $\tilde{\mathbf{U}}^{r}(t)$.

%---------------------------------------------------------------
\begin{figure}[t]
\includegraphics[width=0.49\textwidth]{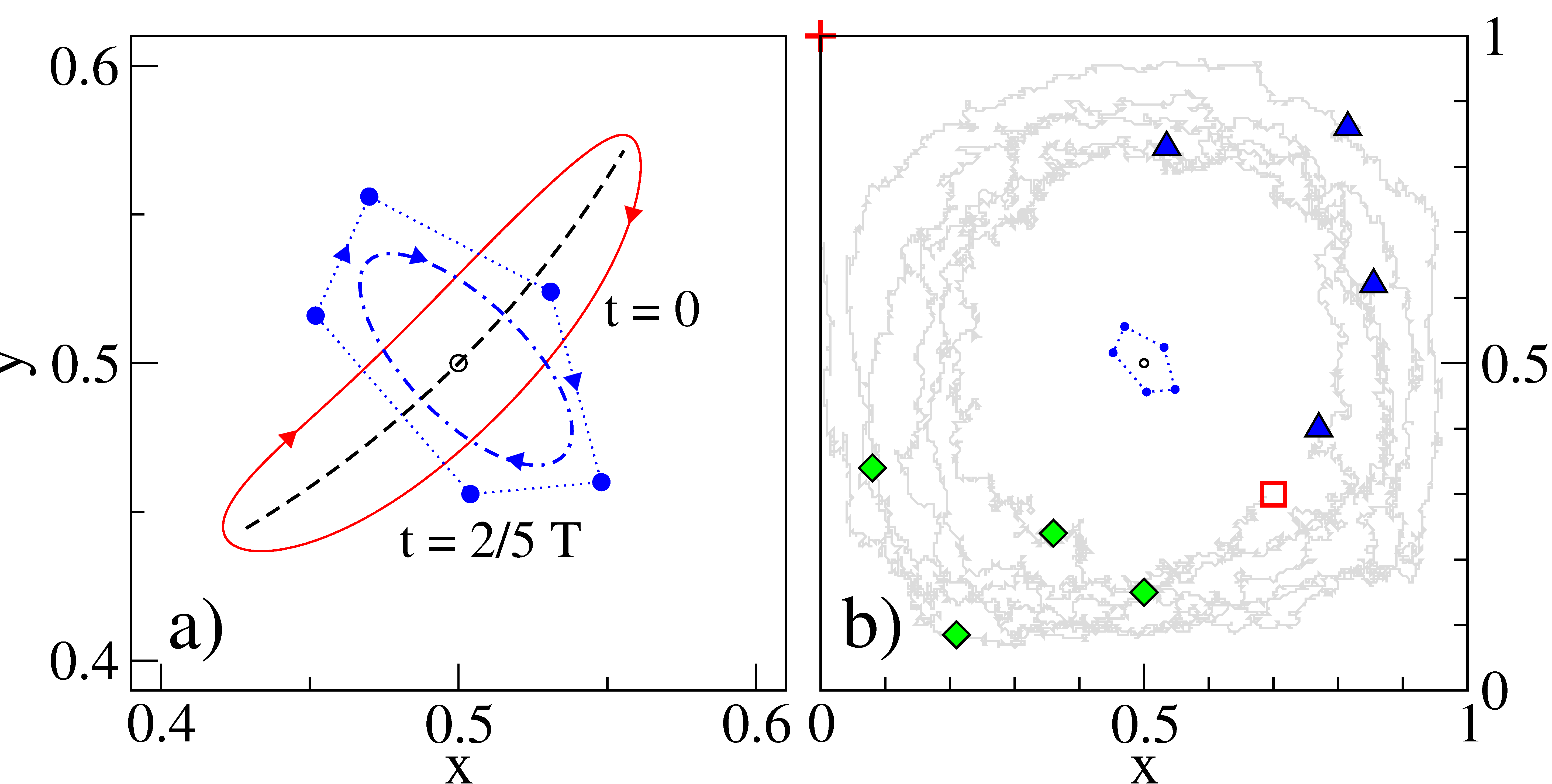}
\caption{(color online) Evolutionary dynamics governed by the BoS game with modulated payoffs. 
(a) Period-one limit cycles of the mean-field dynamics for 
 $\omega = 0.1$ (blue dash-dotted  line) 
and $\omega = 0.01$ (red solid  line) are localized near the Nash equilibrium  of the stationary game,
$\left(\frac{1}{2},\frac{1}{2}\right)$ 
(arrows indicate the direction of motion). In the limit $\omega \rightarrow 0$, the mean-field attractor 
shrinks to the set of adiabatic Nash equlibria (black dashed line). 
Mean position ({\color{blue}$\bullet$}) of a finite-$N$ metastable 
Floquet state, $\left(\bar{x}(t), \bar{y}(t)\right)$, Eq.~(\ref{Eq:mean}), moves along the limit cycle localized 
near the point $\left(\frac{1}{2},\frac{1}{2}\right)$ (the means are plotted at the instants
$t_n = nT/5$, $n=0,..,4$); (b) A stochastic trajectory (grey line) 
reveals the existence of a period-two limit cycle [the period doubling can be resolved with stroboscopic 
points, plotted at the instants $2nT$ ($\bigtriangleup$) and $(2n+1)T$ ($\diamondsuit$)]. The 
trajectory is initiated at the point marked with the open blue square and ends up at the absorbing state (red cross at 
the upper left corner). The trajectory of the mean of the finite-$N$ metastable 
Floquet state ({\color{blue}$\bullet$})  is distinctively different from the stochastic trajectory 
[note the change of scale as compared to panel (a)].
The parameters  are $f = 0.5$, $N=200$, and $M=T/\triangle t=10N$ 
(corresponds to the driving frequency  $\omega=0.1$ in the mean-field limit) \cite{size_issue}. 
Other parameters as in Fig. 2.}\label{Fig:3}
\end{figure}
%---------------------------------------------------------------

The evolution of the means of the pdf $\mathbf{d}(t)$ (see Fig.~4a), $\left(\bar{x}(t), \bar{y}(t)\right)$, 
is close to the period-one limit cycle, 
see  blue dots on Fig.~3a. However, the Floquet state consists of two peaks 
produced by the noised period-two limit cycle (compare also the positions of the  stroboscopic points in 
Fig.~3b with the pdf
for $t=0$ in Fig.~4a). The peak contributions balance each other thus reducing the 
dynamics of the means to the vicinity of the the point $\left(\frac{1}{2},\frac{1}{2}\right)$. 
%---------------------------------------------------------------
\begin{figure}[t]
\includegraphics[width=0.45\textwidth]{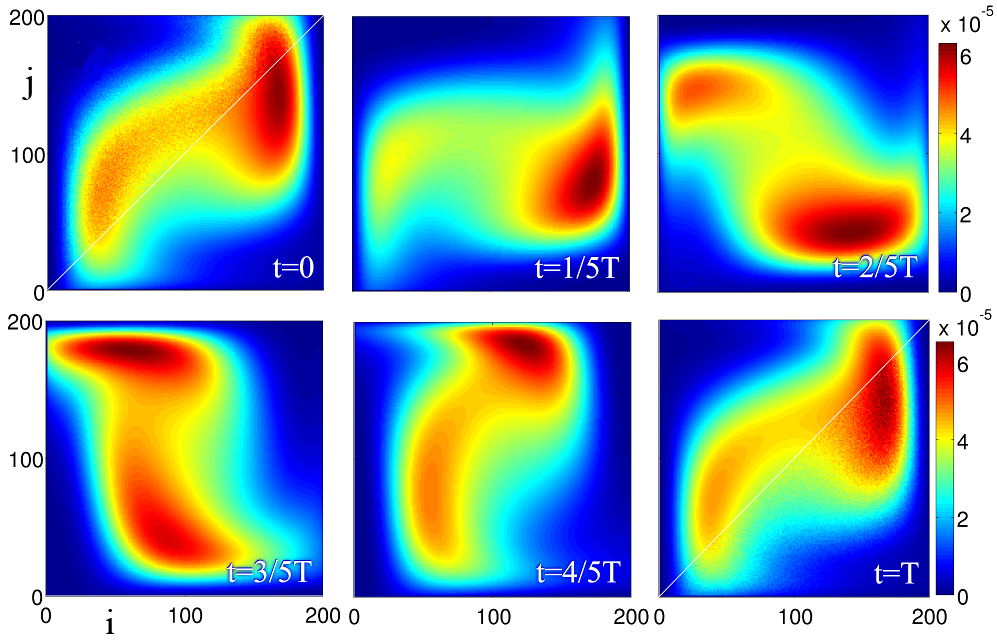}
\caption{(color online) Evolution of the metastable Floquet state over one period of modulations. 
The pdfs obtained by the direct diagonalization 
of the reduced period-one propagator for $N=200$. The corresponding means $\left(\bar{x}(t), \bar{y}(t)\right)$ 
are shown on Fig. 3a ({\color{blue} $\bullet$}). Plots for $t = 0$ (above the diagonal) and $t = T$ (below the diagonal)
present the results of the  stochastic sampling. 
%(b) The  lifetime $t_{\mathrm{life}}$ as a function of the modulation strength $f$, 
%for the population size $N = 50$ ($\bigcirc$), $100$ ($\Box$), and $200$ ($\triangle$). 
%Other parameters are as in Fig. 2.             
}\label{Fig:4}
\end{figure}
%---------------------------------------------------------------
The lifetime of the state $\mathbf{d}(t)$ can be estimated with the largest eigenvalue $\lambda_T$, $0<\lambda_T<1$, 
of the  matrix $\tilde{\mathbf{U}}^{r}(T)$. To compare it with 
lifetimes of stationary metastable states, we introduce the mean single-round 
exponent, $\bar{\lambda}_T = \lambda_T^{1/M}$ and define the mean 
lifetime as $t_{\mathrm{life}} = 1/(1-\bar{\lambda}_T)$ \cite{generic}. 
%Figure 4b shows the dependence of $t_{\mathrm{life}}$ on the strength of modulations. 
Aside of the slow decay trend, we found the effect of modulations not being strong. This is in stark contrast 
to the structure of the metastable states. Namely, while in 
the stationary limit the pdf $\mathbf{d}$ is localized near the Nash equilibrium, 
at the maximal distance from the absorbing boundaries, 
the metastable Floquet state is localized near the absorbing boundary, see Fig. 4. 
We also detect the increase of the boundary localization with the increase of the population 
size beyond $N = 200$. This suggests that, in the  limit $N \rightarrow \infty$, the dynamics 
of the system is governed by a period-two limit cycle localized near the absorbing boundary. 
The boundary localization of the metastable attractor can be interpreted
as the presence of small fractions of mutants
\cite{smith}, i.e. the players that are using strategies different from that used by the 
majority of populations. The evolutionary dynamics of the mutant fractions looks like a repeating 
sequence of population bottlenecks \cite{dawkins,bottleneck} 
yet this only weakly affects fraction lifetimes \cite{issue1} even in the case of finite $N$.

\textit{Conclusions.} We presented a concept of metastable Floquet 
states in game-driven populations when mate selection preferences are periodically 
changing in time. Here we combined the Floquet formalism with 
the concept of quasi-stationary distributions to reveal 
the existence of complex, liquid-like nonequlibrium 
dynamics in the strategy space which cannot be resolved within the mean-field framework.
%By combining the Floquet ideology with the concept of quasi-stationary
%distributions, we reveal the existence of complex nonequlibrium  dynamics which cannot be resolved within
%the mean-field framework.
Metastable Floquet states are not restricted to the field of ecology studies
but can emerge in different periodically modulated systems with stochastic event-driven dynamics. 
They may, for example, underlay a gene expression in a single cell, 
which is modulated by a circadian rhythm \cite{cyrcadian}
and can provide new interpretations of the Bose-Einstein condensation in ac-driven
atomic ensembles \cite{ketz,frey2}.
%%%%%%%%%%%%%%%%%%%%%%%%%%%%%%%%%%%%%%%%%%%%%

%%%%%%%%%%%%%%%%%%%%%%%%%%%%%%%%%%%%%%%%%%%%%%%
\begin{center}
\textbf{Supplemental Material}
\end{center}

\section{Seasonal variations in mate preferences}
Stable time variations were found in the  female flycatcher preferences 
or male forehead patch size that resulted in late-breeding females 
preferring males with larger patches \cite{fly}. It was explained by 
the fact that in the beginning of the breeding season, large-patched 
males allocate more resources to courting than to parental care but 
change their habits to the opposite late in the season. Seasonal 
variations were also found in fiddle crabs (female  preference to 
male claw size) \cite{crab}, two-spotted goby (female  preference 
to overall male size) \cite{goby}, and sailfin mollies (male preferences 
for two different kind of females) \cite{molly}. 

%There is overall no 
%agreement reached between ecologists on the role the seasonal plasticity 
%in the mate preference (of different sexually selected traits) plays in 
%the determination of the evolution direction of a species, see Ref. \cite{srev}.

\section{Moran process}
Players $A$ (males) and $B$ (females) form two 
populations, each one of a fixed size $N$ and with two available strategies, $s = \{1,2\}$. 
Payoffs are specified by four functions, $\{a_{ss'}(t)\}$ and $\{b_{s's}(t)\}$, $s,s' = \{1,2\}$. 
The  average payoff of the players using strategy $s$ is
\begin{eqnarray}
~\pi^{A}_s(j,t) = a_{s1}(t)\frac{j}{N}+a_{s2}(t)\frac{(N-j)}{N}, \\
\pi^{B}_s(i,t) = b_{s1}(t)\frac{i}{N}+b_{s2}(t)\frac{(N-i)}{N}.~
\label{Eq:quasi_sym1}
\end{eqnarray}
Payoffs determine the probabilities for a player to be chosen for reproduction, e.g. for the male population, 
\begin{eqnarray}
P^{A}_s(i,j,t) = \frac{1}{N} \cdot \frac{1 - w + w\pi^{A}_s(j,t)}{1 -w + w \bar{\pi}^{A}(i,j,t)},
\label{Eq:quasi_sym2}
\end{eqnarray}
where $\bar{\pi}^{A}(i,j,t) = [i\pi^{A}_{1}(j,t)+(N-i)\pi^{A}_{2}(j,t)]/N$ is the average payoff 
of the males. The baseline fitness $w \in [0,1]$ is a  tunable baseline 
fitness parameter determining how the player's chance to be chosen for reproduction  
is related to player's performance \cite{moranNature,traulsen1}. 
When $w = 0$, the probability to be chosen for reproduction 
does not depend on player's performance and is uniform across the population.

After the choice has been 
made, another member of the population is chosen completely randomly and replaced with an 
offspring of the player chosen for reproduction, i.e. with a player using the same 
strategy as its parent \cite{imitation}. This update mechanism is acting simultaneously 
in both populations, $A$ and $B$, such that a mating pair produces two offspring, 
a male and a female, on every round. Therefore, the size of the populations $N$ 
remains constant.
%preserved and $N$ is a game parameter.

A single round can be considered as a one-step Markov process, 
with transition rates, e.g. for population $A$, from a state $i$
to states $i+1$ and $i-1$, are given by \cite{traulsen1, traulsen4}
\begin{eqnarray} \nonumber
T_{A}^{+}(i,j,t) =\frac{1-w + w\pi^{A}_1(t)}{1 -w + w \bar{\pi}^{A}}\frac{i}{N}\frac{N-i}{N}, \\
T_{A}^{-}(i,j,t) =\frac{1-w + w\pi^{A}_2(t)}{1 -w + w \bar{\pi}^{A}}\frac{N-i}{N}\frac{i}{N}.
\label{Eq:rates}
\end{eqnarray}

\section{Transition tensor}
Here we describe the transition fourth-order tensor $S^{m}(i,j,i',j')$ 
in terms of the rates [$T_{A}^{+,-}(i,j,t)$ and $T_{B}^{+,-}(i,j,t)$] 
for populations $A$ and $B$ given by Eq.~(4) in the main text. 
The stochastic Moran process can be expressed as a Markov chain \cite{traulsen4}
\begin{widetext}
\begin{eqnarray}
p^{m+1}(i,j) &=& \left[1-T_{A}^{+}(i,j,m\triangle t)-T_A^{-}(i,j,m\triangle t)\right]\left[1-T_B^{+}(i,j,m\triangle t)-T_B^{-}(i,j,m\triangle t)\right]p^{m}(i,j) \nonumber \\
&& + T_B^{-}(i,j+1,m\triangle t)\left[1-T_A^{-}(i,j+1,m\triangle t)-T_A^{+}(i,j+1,m\triangle t)\right]p^{m}(i,j+1) \nonumber \\
&& + T_B^{+}(i,j-1,m\triangle t)\left[1-T_A^{-}(i,j-1,m\triangle t)-T_A^{+}(i,j-1,m\triangle t)\right]p^{m}(i,j-1) \nonumber \\
&& + T_A^{-}(i+1,j,m\triangle t)\left[1-T_B^{-}(i+1,j,m\triangle t)-T_B^{+}(i+1,j,m\triangle t)\right]p^{m}(i+1,j) \nonumber \\
&& + T_A^{+}(i-1,j,m\triangle t)\left[1-T_B^{-}(i-1,j,m\triangle t)-T_B^{+}(i-1,j,m\triangle t)\right]p^{m}(i-1,j) \nonumber \\ 
&& + T_A^{-}(i+1,j+1,m\triangle t)T_B^{-}(i+1,j+1,m\triangle t) p^{m}(i+1,j+1) \nonumber \\ 
&& + T_A^{+}(i-1,j+1,m\triangle t)T_B^{-}(i-1,j+1,m\triangle t) p^{m}(i-1,j+1) \nonumber \\
&& + T_A^{-}(i+1,j-1,m\triangle t)T_B^{+}(i+1,j-1,m\triangle t) p^{m}(i+1,j-1) \nonumber \\
&& + T_A^{+}(i-1,j-1,m\triangle t)T_B^{+}(i-1,j-1,m\triangle t) p^{m}(i-1,j-1).
\end{eqnarray}
\end{widetext}

The above equation can be recast into
\begin{eqnarray}
p^{m+1}(i,j) &=& \sum_{i',j'} S^{m}(i,j,i',j') p^{m}(i',j'), 
\end{eqnarray}
where the fourth-order tensor $S^{m}(i,j,i',j')$ is given by,
\begin{widetext}
\begin{eqnarray}
S^{m}(i,j,i',j') &=& \left[1-T_A^{+}(i',j',m\triangle t)-T_A^{-}(i',j',m\triangle t)\right]\left[1-T_B^{+}(i',j',m\triangle t)-T_B^{-}(i',j',m\triangle t)\right]\delta_{i',i}\,\delta_{j',j} \nonumber \\
&& + T_B^{-}(i',j',m\triangle t)\left[1-T_A^{-}(i',j',m\triangle t)-T_A^{+}(i',j',m\triangle t)\right]\delta_{i',i}\,\delta_{j',j+1} \nonumber \\
&& + T_B^{+}(i',j',m\triangle t)\left[1-T_A^{-}(i',j',m\triangle t)-T_A^{+}(i',j',m\triangle t)\right]\delta_{i',i}\,\delta_{j',j-1} \nonumber \\
&& + T_A^{-}(i',j',m\triangle t)\left[1-T_B^{-}(i',j',m\triangle t)-T_B^{+}(i',j',m\triangle t)\right]\delta_{i',i+1}\,\delta_{j',j} \nonumber \\
&& + T_A^{+}(i',j',m\triangle t)\left[1-T_B^{-}(i',j',m\triangle t)-T_B^{+}(i',j',m\triangle t)\right]\delta_{i',i-1}\,\delta_{j',j} \nonumber \\ 
&& + T_A^{-}(i',j',m\triangle t)T_B^{-}(i',j',m\triangle t)\delta_{i',i+1}\,\delta_{j',j+1} \nonumber \\ 
&& + T_A^{+}(i',j',m\triangle t)T_B^{-}(i',j',m\triangle t)\delta_{i',i-1}\,\delta_{j',j+1} \nonumber \\
&& + T_A^{-}(i',j',m\triangle t)T_B^{+}(i',j',m\triangle t)\delta_{i',i+1}\,\delta_{j',j-1} \nonumber \\
&& + T_A^{+}(i',j',m\triangle t)T_B^{+}(i',j',m\triangle t)\delta_{i',i-1}\,\delta_{j',j-1}.
\end{eqnarray}
\end{widetext}
Above $i= 0, \cdots, N$, $j= 0, \cdots, N$, $i'= 0, \cdots, N$, and $j'= 0, \cdots, N$. 
Using the bijection  $k=(N-1)j+i$ and $l=(N-1)j'+i'$, 
we obtain the required matrix form, see Eq.~(7) in the main text.

\section{Adjusted replicator equations}
In the continuous limit $N \rightarrow \infty$, the dynamics of the variables 
$x = i/N$ and $y = j/N$ is defined by the adjusted replicator equations \cite{smith,hof2},
\begin{align} 
&\dot{x} = [1-x][\Delta^A(t) - \Sigma^A(t)y]\frac{1}{\Gamma + \bar{\pi}^A(x,y,t)}, \\
&\dot{y} = [1-y][\Delta^B(t) - \Sigma^B(t)x]\frac{1}{\Gamma + \bar{\pi}^B(x,y,t)},
\label{Eq:replicator}
\end{align}
where $\Delta^{C} = c_{12} - c_{22}$, $\Sigma^{C}=c_{11}+c_{22}-c_{12}-c_{21}$, 
$\Gamma = \frac{1-w}{w}$, and $C=\{A,B\}$.  $\bar{\pi}^A(x,y,t)$ [$ \bar{\pi}^B(x,y,t)$]
is the averaged (over the population) payoff of the males [females].

\section{The lifetime of a metastable state}
The lifetime of the state $\mathbf{d}(t)$ can be estimated with the largest eigenvalue $\lambda_T$, $0<\lambda_T<1$, 
of the  matrix $\tilde{\mathbf{U}}^{r}(T)$. To compare it with 
lifetimes of stationary metastable states, we introduce the mean single-round 
exponent, $\bar{\lambda}_T = \lambda_T^{1/M}$ and define the mean 
lifetime as $t_{\mathrm{life}} = 1/(1-\bar{\lambda}_T)$. 
Figure 1 shows the dependence of $t_{\mathrm{life}}$ on the strength of modulations. 
%---------------------------------------------------------------
\begin{figure}[t]
\includegraphics[width=0.45\textwidth]{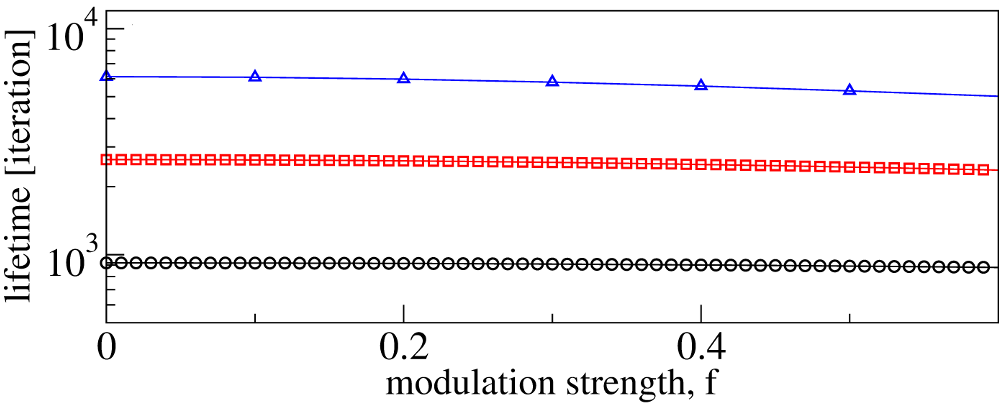}
\caption{(color online) The  lifetime $t_{\mathrm{life}}$ as a function of the modulation strength $f$, 
for the population size $N = 50$ ($\bigcirc$), $100$ ($\Box$), and $200$ ($\triangle$). 
Other parameters are as in Fig. 2 in the main text.
}\label{Fig:1}
\end{figure}
%---------------------------------------------------------------

\section{Simulations}

The preconditioned stochastic sampling  was performed by 
launching trajectories from random initial points, 
uniformly distributed on the $N-1 \times N-1$ 
grid and then sampling the pdf with only those trajectories which remained unabsorbed after $10\cdot N^2$ rounds.

For $N=200$ the diagonalization of 
the $39~ 601 \times 39~ 601$ matrix $\tilde{\mathbf{Q}}_T$ was performed 
on  the cluster of the MPIPKS (Dresden) and Leibniz-Rechenzentrum (M\"{u}nchen).
The stochastic  sampling was performed on a GPU cluster consisting of twelve TESLA K20XM cards. 
That allowed us to obtain
$5 \cdot 10^8$ realizations for each set of parameters.

%For $N=200$ we find a perfect agreement between the results of the two approaches. 


\begin{thebibliography}{1000}

\bibitem{sexual} M. Andersson, \textit{Sexual Selection}(Princeton Univ. Press. Princeton, 1994).

\bibitem{dawkins} R. Dawkins, \textit{The Selfish Gene} (Oxford University Press, Oxford, 1976).

\bibitem{generic}  See Supplemental Material for more information.

\bibitem{lizard1} D. Crews, Science \textbf{189}, 1059 (1975).

\bibitem{lizard2} M. M. Holmes, C. L. Bartrem, and J. Wade,  Physiol. and Behav. \textbf{91}, 601 (2007).

\bibitem{srev} V. D. Jennions and M. Petrie, Biol. Rev. Cambridge Philos Soc., \textbf{72}
(2006).

\bibitem{darroch}  J. N. Darroch and E. Seneta, J. Appl. Prob. \textbf{2}, 88 (1965).

\bibitem{floquet}  G. Floquet, Annales de l'\'{E}cole Normale Sup\'{e}rieure \textbf{12}, 47 (1883).

\bibitem{floquetH} M. Grifoni and P. H\"{a}nggi, Phys. Rep. \textbf{304}, 229 (1998).


\bibitem{meta1}  G. Biroli and J. Kurchan,  Phys. Rev. E \textbf{64}, 016101 (2001).

\bibitem{meta2}  S. Rulands, T. Reichenbach, and E. Frey, J. Stat. Mech. L01003 (2011).

\bibitem{assaf}  M. Assaf and M. Mobilia,  Phys. Rev. Lett. \textbf{109}, 188701 (2012).


\bibitem{hof2}
J. Hofbauer and K. H. Schlag, J. of Evol. Economics \textbf{10}, 523 (2000).

\bibitem{schlag} K. H. Schlag, J. of Econom. Theory \textbf{78}, 130.

\bibitem{traulsen1} A. Traulsen, J. C. Claussen, and C. Hauert, Phys. Rev. Lett. \textbf{95}, 238701 (2005).

\bibitem{traulsen2}  Ch. S. Gokhale and A. Traulsen,  Dyn. Games and Appl. \textbf{4}, 468 (2014).

\bibitem{traulsen3}  A. Traulsen, J. C. Claussen, and C. Hauert, Phys. Rev. E. \textbf{74}, 011901 (2006).

\bibitem{frey}  A. Dobrinevski and E. Frey, Phys. Rev. E \textbf{85}, 051903 (2012).

%\bibitem{pom}  A. Pomiankowski, J. Theor. Biol. 128, 195 (1987).



\bibitem{moran}  P. A. P. Moran, \textit{The Statistical Processes of Evolutionary Theory} (Clarendon, Oxford, 1962).

\bibitem{moranNature}  M. A. Nowak, A. Sasaki, C. Taylor, and D. Fudenberg, Nature \textbf{428},  646 (2004).

%\bibitem{imitation} These two consecutive steps, death and birth, can 
%be  reinterpreted as a single step of \textit{imitation}, i.e. adoption of the strategy of the 
%first player by the second one \cite{hof2,schlag}.


\bibitem{smith} J. M. Smith, \textit{Evolution and the Theory of Games}(Cambridge University Press, Cambridge, 1982).

%\bibitem{transition}  Here we follow the  convention that the stochastic matrix acts on 
%the probability column vector to the right. 


\bibitem{gant}  E. Seneta, \textit{Non-negative Matrices and Markov Chains} (Springer, NY, 2006).

\bibitem{dykman}  M. Khasin and M. I. Dykman, Phys. Rev. Lett. \textbf{103}, 068101 (2009).

\bibitem{FPT} By virtue of the Perron-Frobenius 
theorem, $\lambda$  and $\tilde{\mathbf{d}}$ are both real and 
non-negative \cite{gant}.

\bibitem{neumann}  This choice corresponds 
to the  Matching Pennies game, see J. von Neumann and O. Morgenstern, 
\textit{Theory of Games and Economic Behaviour} (Princeton University Press, Princeton, 1944).


\bibitem{nash} J. Nash, PNAS \textbf{36}, 48 (1950).

\bibitem{traulsen4} A. Trauslen, J. C. Claussen, and C. Hauert, Phys. Rev. E \textbf{85}, 041901 (2012).

\bibitem{arnold} L. Arnold, \textit{Random Dynamical Systems} (Springer, NY, 2003).

\bibitem{lindner}  P. J. Thomas and B. Lindner,  Phys. Rev. Lett. \textbf{113}, 254101 (2014).

%\bibitem{redner} S. Redner, \textit{A Guide to First Passage Processes} 
%(Cambridge University Press, Cambridge, 2001).

\bibitem{traulsen5} A. J. Black, A. Traulsen, and T. Galla, Phys. Rev. Lett. \textbf{109}, 028101 (2012).


\bibitem{meantime} The average absorption time for a specific initial 
state, $t_{\mathrm{abs}}(i,j)$,
is proportional the corresponding entry in the \textit{left} maximal-eigenvalue eigenvector 
of the reduced matrix $\tilde{\mathbf{Q}}$. The proportionality coefficient can be
found from the dual orthonormality condition. 




\bibitem{size_issue} 
We find a sharp contrast between the mean-filed dynamics 
and the stochastic evolution for this particular value of  $\omega$. 
The optimal value for the frequency (period) of modulations could be different
for other driving scheme and/or other choice of the game payoffs.

\bibitem{bottleneck} T. Maruyama and P. A. Fuerst, Genetics \textbf{111}, 691 (1985).

\bibitem{issue1}  The relations between 
the exponent $\lambda_T$,
mean absorption (fixation) time \cite{meantime}, and dynamical properties of Floquet states is an 
interesting issue. It can be explored, for example, with a 
discrete-time generalization of the ``optimal path to exctintion'' approach \cite{dykman2,escudero,meerson}.

\bibitem{dykman2}  M. I. Dykman, E. Mori, J. Ross, and P. M. Hunt, J. Chem. Phys. \textbf{100}, 5735 (1994).

\bibitem{escudero} C. Escudero and J. A. Rodriguez, Phys. Rev. E \textbf{77}, 011130 (2008).

\bibitem{meerson}   M. Assaf, A. Kamenev, and B. Meerson, Phys. Rev. E \textbf{78}, 041123 (2008).

%\bibitem{floquet_ecol1} C. A. Klausmeier, Theor. Ecol. \textbf{1}, 153 (2008); 
%C. T. Kremer and C. A. Klausmeier, J. Theor. Biol. \textbf{339}, 14 (2013).

\bibitem{cyrcadian} 
S. S. Golden, V. M. Cassone, and A. Li Wang, Nat. Struct.  Mol. Biol. \textbf{14}, 362 (2007);
A. Sancar, Nat. Struct.  Mol. Biol. \textbf{15}, 23  (2008).

\bibitem{ketz} D. Vorberg, W. Wustmann, R. Ketzmerick, and A. Eckardt, Phys. Rev. Lett. \textbf{111}, 240405 (2013).

\bibitem{frey2} J. Knebel, M. F. Weber, T. Kr\"{u}ger, and E. Frey, Nature Comm. \textbf{6}, 6977 (2015). 




\end{thebibliography}

\begin{thebibliography}{1000}
\bibitem{fly} A. Qvarnstr\"{o}m, T. P\"{a}rt, and B. C. Sheldon, Nature \textbf{405}, 344 (2000).
\bibitem{crab} R. N. C. Milner, \textit{et al.}, Behav. Ecology \textbf{21}, 311(2010).
\bibitem{goby} A. A. Borg, E. Forsgren, and T. Amudsen, Anim. Behav. \textbf{72}, 763 (2006).
\bibitem{molly} K. U. Heubel and J. Schlupp, Behav. Ecol. 19, 1080 (2008).
%\bibitem{srev} V. D. Jennions and M. Petrie, Biol. Rev. Cambridge Philos Soc., \textbf{72}
%(2006).



%\bibitem{moran}  P. A. P. Moran, \textit{The Statistical Processes of Evolutionary Theory} (Clarendon, Oxford, 1962).

\bibitem{moranNature}  M. A. Nowak, A. Sasaki, C. Taylor, and D. Fudenberg, Nature \textbf{428},  646 (2004).

\bibitem{traulsen1} A. Traulsen, J. C. Claussen, and C. Hauert, Phys. Rev. Lett. \textbf{95}, 238701 (2005).

\bibitem{imitation} These two consecutive steps, death and birth, can 
be  reinterpreted as a single step of \textit{imitation}, i.e. adoption of the strategy of the 
first player by the second one \cite{hof2,schlag}.

\bibitem{hof2}
J. Hofbauer and K. H. Schlag, J. of Evol. Economics \textbf{10}, 523 (2000).

\bibitem{schlag} K. H. Schlag, J. of Econom. Theory \textbf{78}, 130.

\bibitem{traulsen4} A. Trauslen, J. C. Claussen, and C. Hauert, Phys. Rev. E \textbf{85}, 041901 (2012).

\bibitem{smith} J. M. Smith, \textit{Evolution and the Theory of Games}(Cambridge University Press, Cambridge, 1982).

%\bibitem{Trauslen} A. Trauslen, J. C. Claussen, and C. Hauert, Phys. Rev. E \textbf{85}, 041901 (2012).
\end{thebibliography}
\end{document}